\documentclass[conference]{IEEEtran}
\IEEEoverridecommandlockouts
\usepackage{cite}
\usepackage{amsmath,amssymb,amsfonts}
\usepackage{algorithmic}
\usepackage{graphicx}
\usepackage{textcomp}
\usepackage{xcolor}
\def\BibTeX{{\rm B\kern-.05em{\sc i\kern-.025em b}\kern-.08em
    T\kern-.1667em\lower.7ex\hbox{E}\kern-.125emX}}
\begin{document}

\title{Interactive Manipulation and Visualization of 3D Brain MRI for Surgical Training}

\author{\IEEEauthorblockN{Zichen Gui}
\IEEEauthorblockA{\textit{Department of Computing Science} \\
\textit{University of Alberta, Edmonton, Canada}\\
Email: zgui@ualberta.ca}
\and
\IEEEauthorblockN{Siddharth Jha}
\IEEEauthorblockA{\textit{Department of Computing Science} \\
\textit{University of Alberta, Edmonton, Canada}\\
Email: sjha4@ualberta.ca}
\and
\IEEEauthorblockN{Benjamin Delbos}
\IEEEauthorblockA{\textit{Department of Electrical Engineering} \\
\textit{INSA Lyon, Lyon, France}\\
Email: benjamin.delbos@insa-lyon.fr}
\and
\IEEEauthorblockN{Richard Moreau}
\IEEEauthorblockA{\textit{Department of Electrical Engineering} \\
\textit{INSA Lyon, Lyon, France}\\
Email: richard.moreau@insa-lyon.fr}
\and
\IEEEauthorblockN{Arnaud Leleve}
\IEEEauthorblockA{\textit{Department of Electrical Engineering} \\
\textit{INSA Lyon, Lyon, France}\\
Email: arnaud.leleve@insa-lyon.fr}
\and
\IEEEauthorblockN{Irene Cheng}
\IEEEauthorblockA{\textit{Department of Computing Science} \\
\textit{University of Alberta, Edmonton, Canada}\\
Email: locheng@ualberta.ca}
}

\maketitle

\begin{abstract}
In modern medical diagnostics, magnetic resonance imaging (MRI) is an important technique that provides detailed insights into anatomical structures. In this paper, we present a comprehensive methodology focusing on streamlining the segmentation, reconstruction, and visualization process of 3D MRI data. Segmentation involves the extraction of anatomical regions with the help of state-of-the-art deep learning algorithms. Then, 3D reconstruction converts segmented data from the previous step into multiple 3D representations. Finally, the visualization stage provides efficient and interactive presentations of both 2D and 3D MRI data. Integrating these three steps, the proposed system is able to augment the interpretability of the anatomical information from MRI scans according to our interviews with doctors. Even though this system was originally designed and implemented as part of human brain haptic feedback simulation for surgeon training, it can also provide experienced medical practitioners with an effective tool for clinical data analysis, surgical planning and other purposes.
\end{abstract}


\section{Introduction}
Medical imaging plays an important role in modern healthcare systems, as it is able to provide clear insights into internal structures. MRI is the most heavily utilized of imaging services. It works by employing radio waves and magnetic fields to capture the inside of the body. Compared with computed tomography (CT), MRI can provide better image contrast especially when dealing with soft tissues \cite{Hashemi_Lisanti_Bradley_2018}. 

However, the legibility of MRI scans can be compromised by the restricted characteristics of 2D greyscale images. In this context, our work contributes by developing a research component to enhance their visual representations. While this research was originally part of haptic feedback simulation, its real-world application can extend far beyond the initial intention and it can be applied for patient data analysis, clinical decision-making, and other purposes.

This research aims to design a framework to streamline the automatic segmentation, reconstruction, and visualization of MRI data to increase its comprehensibility. The main objectives of the proposed system are to allow easy MRI file selection, utilize state-of-the-art deep learning algorithms to segment the selected MRI file, develop a workflow to reconstruct segmentation data into 3D representations, implement a user-friendly interface for visualization of the outcome, integrate these feature into one standalone application with minimal user intervention, and enable high extensibility to allow additional features to build upon for other purposes.

The rest of this paper is structured as follows: Section II summarizes the literature review of related work. Section III presents our methodology for efficient segmentation, reconstruction, and visualization techniques. Section IV discusses the results, followed by doctors' feedback. Finally, Section V concludes our research and proposes potential future work.

\section{Literature Review}

\subsection{Segmentation}

MRI segmentation refers to the technique of drawing boundaries between substructures within the MRI image \cite{Despotović2015}, which plays a crucial role in medical imaging analysis. 

The field of MRI segmentation has historically been dominated by traditional methods such as thresholding. Simple thresholding assigns a defined value to pixels in each region \cite{8058355} \cite{sujji2013mri}. The more adaptive thresholding approach is Otsu’s thresholding, which calculates the threshold value by minimizing the image intra-class variance \cite{Nyo2022}. Atlas-based and surface-based techniques can be used for MRI segmentation as well \cite{Despotović2015}.

However, traditional methods often require MRI scans to follow certain modality and resolution specifications, which limits their flexibility. In recent years, deep learning has significantly impacted this field and featured more reliable results \cite{brainsci11081055}. For example, various convolutional neural networks (CNNs) based models with patch-wise, semantic-wise, or cascaded architectures have been proposed to contribute to MRI segmentation \cite{Akkus2017}. Among these advancements, SynthSeg stands out by its ability to handle variations in contrast and resolution by utilizing a generative network based on the Bayesian segmentation and domain randomization strategy \cite{billot_synthseg_2023} \cite{billot_robust_2023}. As an unsupervised model, SynthSeg achieves a similar level of accuracy as other supervised CNN models and outperforms other state-of-the-art adaptations without retraining, making it suitable for plug-and-play clinical applications.

\subsection{Reconstruction}

3D reconstruction is the process of creating 3D representations from 2D scans by extracting triangle topology and closing the surfaces. One classic 3D reconstruction technique is Marching Cubes, which divides the 3D volumetric space into small cubes and evaluates the scalar field's values at each cube's corners \cite{10.1145/37401.37422}. This determines the most appropriate configuration of its neighbouring voxels within the cube and creates corresponding triangles representing the surfaces. 

On the other hand, Flying Edges \cite{7348069} is an improved 3D reconstruction technique based on Marching Cubes. Unlike standard Marching Cubes which adopts a cube-like approach, Flying Edges works by traversing the grid row by row. It directly analyzes edge information within the scalar field and connects the edges across voxels to create surfaces. When dealing with noisy data, Flying Edges can produce better and faster results than Marching Cubes.

By leveraging the Flying Edges algorithm as well as other optimization techniques such as Taubin smoothing, optional decimation and surface normal computation, 3D Slicer \cite{FEDOROV20121323} has the ability to create a closed surface representation to achieve 3D reconstruction. As a renowned medical image processing tool, 3D Slicer also provides other powerful functions to facilitate the comprehension of various medical data.

\section{Methodology}
\subsection{Segmentation}
MRI segmentation is the foundational step in many medical imaging applications. The precision in identifying and segmenting anatomical structures affects the accuracy of the subsequent stages. In this context, SynthSeg, the state-of-the-art deep learning algorithm, offers several key advantages, making it a compelling choice for this scenario. 

Our work seeks to ensure maximum usability by directly supporting unprocessed MRI scans acquired from medical centres. Therefore, variations in imaging protocols and resolutions in real-world clinical settings are taken care of in our design. While traditional methods often have limited flexibility and their resulting qualities fluctuate, deep learning models like SynthSeg demonstrate better adaptability. Another advantage of SynthSeg lies in its minimal retraining requirement. Although traditional CNN models may have the capability to handle varying data formats to some extent, extensive retraining is required for optimized results. However, SynthSeg's synthesis-based approach effectively bridges the gap between different datasets without the need for frequent adjustments. 


Admittedly, while SynthSeg presents significant advantages, it might take a longer processing time. We have found that it takes several minutes to segment an MRI scan, which is longer than some traditional methods that generate results almost instantly. However, it is also important to emphasize that this computational cost is a one-time occurrence for each new MRI scan, as future viewing of the same MRI scan will not need the execution of SynthSeg. 



Our segmentation process begins with resampling the original MRI file to have a consistent 1mm isotropic resolution, optimized for segmentation and future visualization. The resampled image then undergoes SynthSeg's deep learning model, which analyzes each voxel within the image. We also choose to force SynthSeg to run on CPU instead of GPU for better device compatibility and fewer dependencies. This generates a segmentation mask, where every voxel is assigned a numerical label corresponding to a specific anatomical structure.



\subsection{Reconstruction}
Subsequent to MRI segmentation, 3D reconstruction decides the shapes of brain structures in the 3D environment, and its accuracy directly impacts the final visualization. In this research, we have selected to use 3D Slicer.

The 3D reconstruction stage also needs to ensure a high adaptability to heterogeneous input data formats in the real world, and thus Flying Edges has significant advantages over Marching Cubes. Furthermore, its faster calculation also contributes to greater efficiency. Therefore, 3D Slicer's integration of Flying Edges and multiple optimization and smoothing techniques makes it a preferable choice for our project.

In our workflow, after obtaining segmented Neuroimaging Informatics Technology Initiative (NIFTI) data, we use 3D Slicer to automate the creation of closed surface representations for each structure and generate 3D STL models. Additionally, we employ a three-step process to obtain a 3D model of the entire skull. We first utilize the voting binary hole-filling image filter to fill cavities within the skull structure. Next, we apply label map smoothing to refine the skull representation, enhanced with anti-aliasing and Gaussian smoothing. Finally, we export the full skull data to a separate STL file and place it in the same folder as the segmented STL files. 


\subsection{Visualization}
Following the previous phases, the dataset now comprises a resampled NIFTI file including volumetric medical scan data at a 1mm isotropic resolution, and multiple STL files representing the 3D surface geometry for the skull and segmented anatomical structures. 

For effective visualization of these data formats, our chosen implementation platform for integration is Three.js. As a robust JavaScript library \cite{Danchilla2012}, Three.js allows for web-based visualization, allowing easy access across devices without installing specific software. Furthermore, Three.js has native support for various 3D model formats such as STL. Thanks to its strong community support, third-party tools such as NIFTI reader \cite{RII_2023} also lower the difficulty of visualizing MRI scans.

To be more specific, our web application includes a 2D MRI viewer and an interactive 3D viewer.

\subsubsection{2D Scan}
2D visualization starts with integrating the open source javascript NIFTI reader \cite{RII_2023} to parse the data. After taking a resampled NIFTI file as input, it outputs the voxel data structured as a 3D array and the image header information, which provides useful information such as dimension and transformation of the NIFTI file. Then, a slicing operation is conducted on the 3D voxel data array to display axial, sagittal, and coronal views of the MRI scan, which involves iterating through each axis—x, y, and z—of the voxel data array. Sliders have also been added to the web page to allow users to control the depth of slicing on each axis. In the last step, the sliced voxel data is mapped and loaded as the texture of plane geometry in Three.js. 

\subsubsection{3D Model}
3D visualization begins with loading the STL files into a Three.js scene using the STLLoader, which retrieves geometry and creates Three.js objects accordingly. Once loaded, these models are grouped to allow for better management and manipulation. 
Three.js's orbit controls are also included in our application so that users can freely navigate the scene and inspect 3D segmentations from different perspectives. Finally, in order to distinguish different structures visually, distinct colours are assigned to each model. Transparency has been adjusted to enhance the visibility of different components in case of overlapping. 

\subsection{Integration}
Previous sections have described a three-step approach to implement our proposed framework. SynthSeg is a Python-based deep learning algorithm and 3D Slice also supports Python integration. On the other hand, Three.js is a browser-based visualization library using HTML and JavaScript. 

Therefore, we use Django \cite{Django_2023}, a Python-based web framework, to streamline the integration process. With the help of Django, HTTP requests (GET and POST) can be used for cross-language communication between JavaScript and Python. In addition, we have employed a lightweight database system, SQLite, to store data associated with each MRI file. In this way, users can see a list of previously accessed files. Introducing a database also prevents redundant segmentation and reconstruction efforts each time the same file is accessed.

\begin{figure}[htbp]
\hspace{0.68cm}
\centerline{\includegraphics[width=0.38\textwidth]{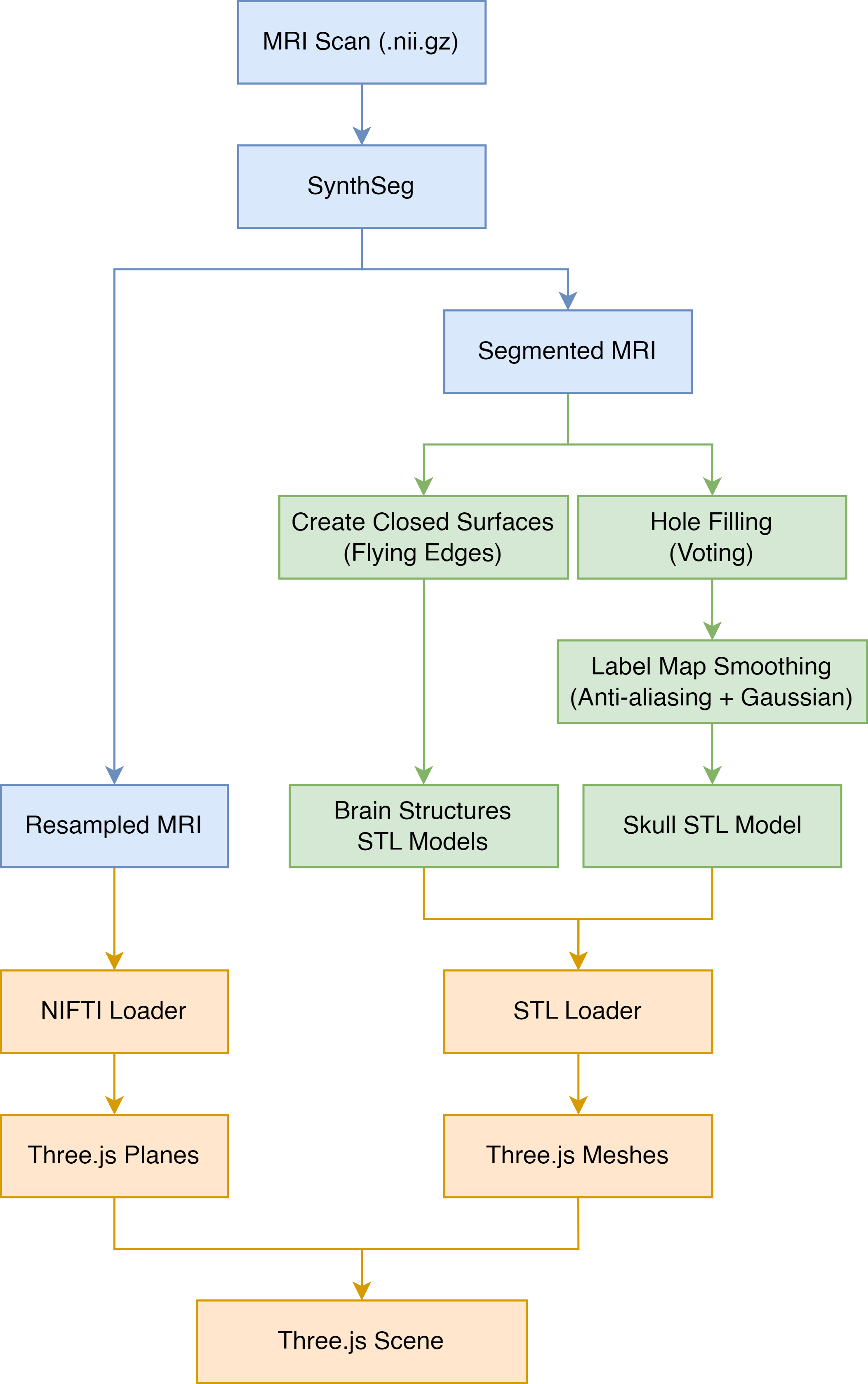}}
\caption{Workflow of the proposed pipeline.}
\label{workflow}
\end{figure}

Fig. \ref{workflow} summarizes the entire workflow of our proposed pipeline, where the blue zone is segmentation, the green zone is reconstruction and the orange zone is visualization.

\section{Results and Discussion}
\subsection{Visual Representations}
During the experiments, we tested our system with multiple MRI files and two of them were selected as samples in this paper. Table \ref{time-tab} shows the time consumed by each step for these two MRI files, running on a test environment with AMD Ryzen 7 6800H and Ubuntu 22.04. Both the first time (when segmentation and reconstruction are needed) and the second time (when the results are reused) data are recorded.

\begin{table}[htbp]
\caption{Table of Workflow Duration for the Selected MRI Files}
\begin{center}
\begin{tabular}{|c|c|c|c|c|}
\hline
\textbf{Filename}&\multicolumn{2}{|c|}{PATIENT\_01.nii.gz} & \multicolumn{2}{|c|}{PATIENT\_05.nii.gz}\\
\hline
\textbf{Resolution}&\multicolumn{2}{|c|}{512 * 512 * 150 mm}&\multicolumn{2}{|c|}{256 * 400 * 400 mm}\\
\hline
\textbf{1$^{st}$ or 2$^{nd}$ Time}& 1$^{st}$& 2$^{nd}$ & 1$^{st}$ & 2$^{nd}$\\
\hline
\textbf{Segmentation}&362.77 s&0 s&321.42 s&0 s\\
\hline
\textbf{Reconstruction}&15.50 s&0 s&13.10 s&0 s\\
\hline
\textbf{Visualization}&1.15 s&1.16 s&2.28 s&1.30 s\\
\hline
\textbf{Total}&379.42 s&1.16 s&336.8 s&1.30 s\\
\hline
\end{tabular}
\label{time-tab}
\end{center}
\end{table}



\begin{figure}[htbp]
\centerline{\includegraphics[width=0.45\textwidth]{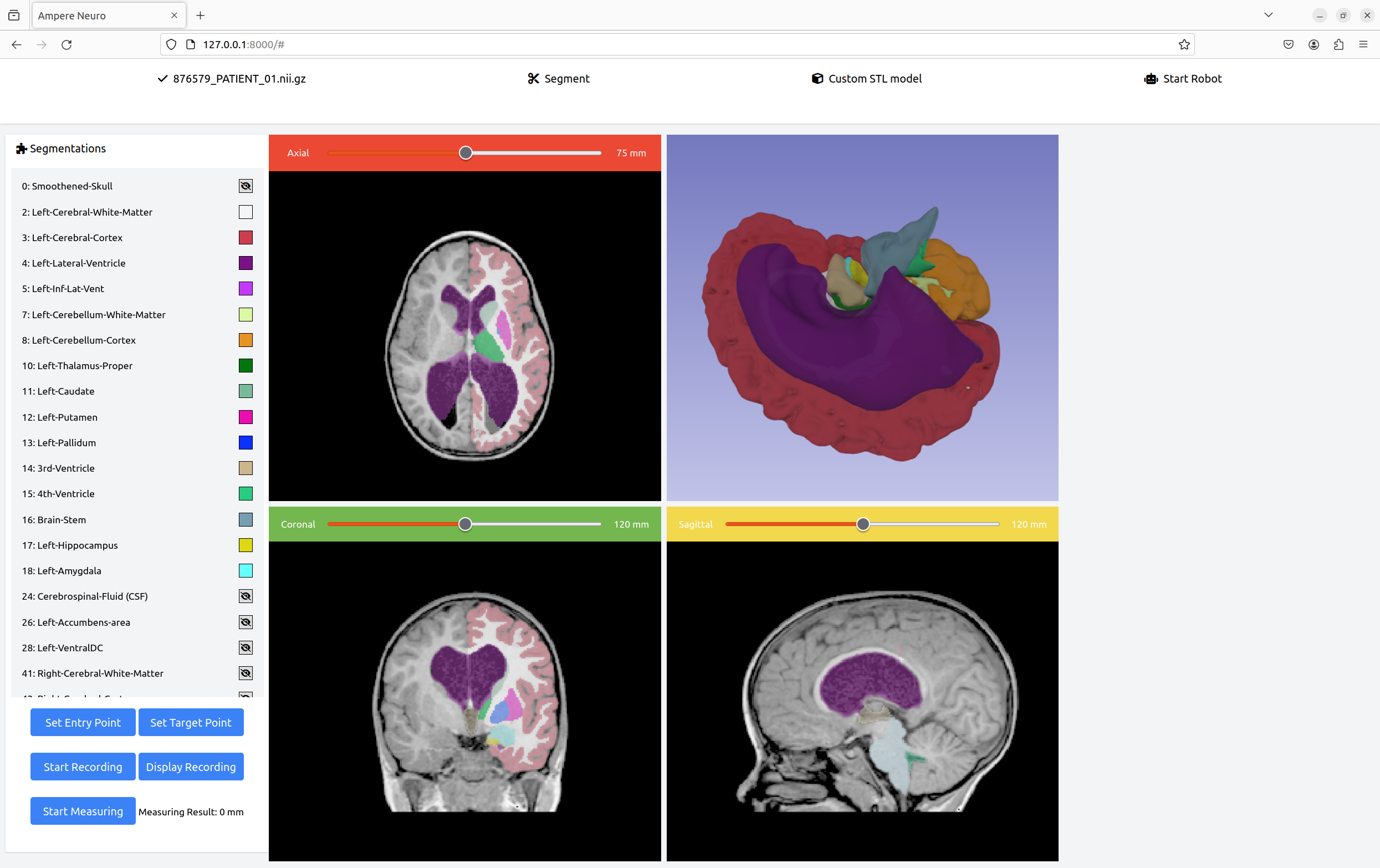}}
\centerline{\includegraphics[width=0.45\textwidth]{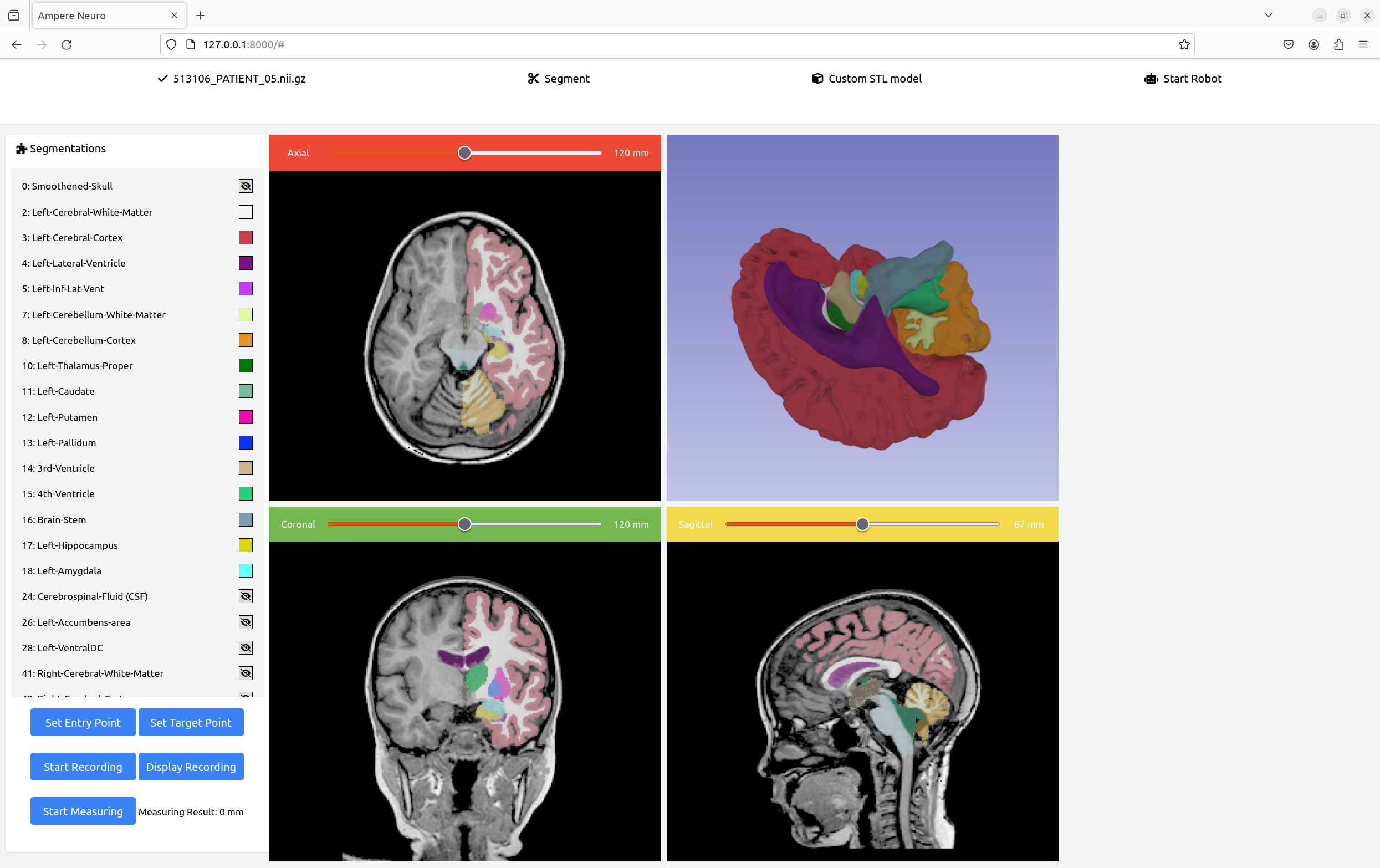}}
\caption{The interface of proposed system displaying with 2D and 3D viewers for MRI visualization.}
\label{screenshot}
\end{figure}

Fig. \ref{screenshot} illustrates the user interface of our system for the two MRI samples. The 2D viewer offers MRI visualization from three perspectives, complete with depth sliders. The 3D viewer enables interactive visualization of the 3D brain structure with different colours. The left panel allows the toggling of a specific brain segmentation and the use of extra tools.

\subsection{Doctor Feedback}
We conducted two remote interviews with experienced doctors to collect their opinions on our framework, each lasting approximately 25 minutes. Each interview started with a complete demonstration of the application using one of the sample MRI files. The doctors were then invited to use the application themselves with the other MRI sample.

After hands-on trials, questions were asked to gather feedback on usability, accuracy, clinical relevance and suggestions.

\subsubsection{Usability}
Both doctors showed positive impressions regarding the graphic interface. The first doctor highlighted the top navigation bar was intuitive. The second noted tooltips could be added when hovering to help first-time users understand each button's function. Both offered similar opinions on the time required for segmentation, stating a few minutes was acceptable but less time would be appreciated.

\subsubsection{Accuracy}
Both doctors acknowledged the application's high accuracy in segmenting anatomical structures. Although they pointed out some errors near the edges of some structures, they also stated that these errors were negligible and were of low risk to decision-making.

\subsubsection{Clinical Relevance}
The first doctor emphasized its application in teaching and training environments, and the second was excited about its potential in surgical planning, especially in collaborative discussions.

\subsubsection{Other Suggestions}
The first doctor recommended we develop a patient-side application by incorporating artificial intelligence techniques that can understand MRI scans and answer related questions. The second advised us to look into adding automatic path planning for surgeries.

\subsection{Discussion}
Insights from the two interviews gave us a practical understanding of the strengths and potential enhancements for our framework. Doctors' positive feedback on usability implied that using Django to integrate the components in our system is an effective choice, resulting in a simplistic workflow. With barely any user intervention, the framework achieved satisfactory outcomes in terms of final visualized brain structures, proving our selection of deep learning algorithms and various tools yield high accuracy. Finally, the doctors' ideas of practical clinical uses, including medical education, surgical planning, and collaborative discussions, displayed the great potential and high extensibility of our proposed system.


\section{Conclusion}
In conclusion, this paper successfully presents a cohesive pipeline that addresses the need for easier interpretation of MRI scans and fulfills the research objectives. 

Looking ahead, several aspects of this project are still open for exploration. Future work can attempt to integrate different deep learning models to further increase accuracy and efficiency. Additionally, its extensibility allows future specialized medical tools to quickly prototype by incorporating it as a base module.



\bibliographystyle{IEEEtran}
\bibliography{cite}

\begin{thebibliography}{10}
\providecommand{\url}[1]{#1}
\csname url@samestyle\endcsname
\providecommand{\newblock}{\relax}
\providecommand{\bibinfo}[2]{#2}
\providecommand{\BIBentrySTDinterwordspacing}{\spaceskip=0pt\relax}
\providecommand{\BIBentryALTinterwordstretchfactor}{4}
\providecommand{\BIBentryALTinterwordspacing}{\spaceskip=\fontdimen2\font plus
\BIBentryALTinterwordstretchfactor\fontdimen3\font minus \fontdimen4\font\relax}
\providecommand{\BIBforeignlanguage}[2]{{%
\expandafter\ifx\csname l@#1\endcsname\relax
\typeout{** WARNING: IEEEtran.bst: No hyphenation pattern has been}%
\typeout{** loaded for the language `#1'. Using the pattern for}%
\typeout{** the default language instead.}%
\else
\language=\csname l@#1\endcsname
\fi
#2}}
\providecommand{\BIBdecl}{\relax}
\BIBdecl

\bibitem{Hashemi_Lisanti_Bradley_2018}
R.~H. Hashemi, C.~J. Lisanti, and W.~G. Bradley, \emph{MRI: The basics}.\hskip 1em plus 0.5em minus 0.4em\relax Wolters Kluwer, 2018.

\bibitem{Despotović2015}
\BIBentryALTinterwordspacing
I.~Despotovi{\'{c}}, B.~Goossens, and W.~Philips, ``Mri segmentation of the human brain: Challenges, methods, and applications,'' \emph{Computational and Mathematical Methods in Medicine}, vol. 2015, p. 450341, Mar 2015. [Online]. Available: \url{https://doi.org/10.1155/2015/450341}
\BIBentrySTDinterwordspacing

\bibitem{8058355}
J.~F. Singh and V.~Magudeeswaran, ``Thresholding based method for segmentation of mri brain images,'' in \emph{2017 International Conference on I-SMAC (IoT in Social, Mobile, Analytics and Cloud) (I-SMAC)}, 2017, pp. 280--283.

\bibitem{sujji2013mri}
G.~E. Sujji, Y.~Lakshmi, and G.~W. Jiji, ``Mri brain image segmentation based on thresholding,'' \emph{International Journal of Advanced Computer Research}, vol.~3, no.~1, p.~97, 2013.

\bibitem{Nyo2022}
\BIBentryALTinterwordspacing
M.~T. Nyo, F.~Mebarek-Oudina, S.~S. Hlaing, and N.~A. Khan, ``Otsu's thresholding technique for mri image brain tumor segmentation,'' \emph{Multimedia Tools and Applications}, vol.~81, no.~30, pp. 43\,837--43\,849, Dec 2022. [Online]. Available: \url{https://doi.org/10.1007/s11042-022-13215-1}
\BIBentrySTDinterwordspacing

\bibitem{brainsci11081055}
\BIBentryALTinterwordspacing
A.~Fawzi, A.~Achuthan, and B.~Belaton, ``Brain image segmentation in recent years: A narrative review,'' \emph{Brain Sciences}, vol.~11, no.~8, 2021. [Online]. Available: \url{https://www.mdpi.com/2076-3425/11/8/1055}
\BIBentrySTDinterwordspacing

\bibitem{Akkus2017}
\BIBentryALTinterwordspacing
Z.~Akkus, A.~Galimzianova, A.~Hoogi, D.~L. Rubin, and B.~J. Erickson, ``Deep learning for brain mri segmentation: State of the art and future directions,'' \emph{Journal of Digital Imaging}, vol.~30, no.~4, pp. 449--459, Aug 2017. [Online]. Available: \url{https://doi.org/10.1007/s10278-017-9983-4}
\BIBentrySTDinterwordspacing

\bibitem{billot_synthseg_2023}
B.~Billot, D.~N. Greve, O.~Puonti, A.~Thielscher, K.~Van~Leemput, B.~Fischl, A.~V. Dalca, and J.~E. Iglesias, ``Synthseg: {Segmentation} of brain {MRI} scans of any contrast and resolution without retraining,'' \emph{{Medical} {Image} {Analysis}}, vol.~86, p. 102789, 2023.

\bibitem{billot_robust_2023}
B.~Billot, Y.~Colin, Magdamo~Cheng, S.~Das, and J.~E. Iglesias, ``{Robust} machine learning segmentation for large-scale analysis of heterogeneous clinical brain {MRI} datasets,'' \emph{{Proceedings} of the {National} {Academy} of {Sciences} ({PNAS})}, vol. 120, no.~9, pp. 1--10, 2023.

\bibitem{10.1145/37401.37422}
\BIBentryALTinterwordspacing
W.~E. Lorensen and H.~E. Cline, ``Marching cubes: A high resolution 3d surface construction algorithm,'' in \emph{Proceedings of the 14th Annual Conference on Computer Graphics and Interactive Techniques}, ser. SIGGRAPH '87.\hskip 1em plus 0.5em minus 0.4em\relax New York, NY, USA: Association for Computing Machinery, 1987, p. 163–169. [Online]. Available: \url{https://doi.org/10.1145/37401.37422}
\BIBentrySTDinterwordspacing

\bibitem{7348069}
W.~Schroeder, R.~Maynard, and B.~Geveci, ``Flying edges: A high-performance scalable isocontouring algorithm,'' in \emph{2015 IEEE 5th Symposium on Large Data Analysis and Visualization (LDAV)}, 2015, pp. 33--40.

\bibitem{FEDOROV20121323}
\BIBentryALTinterwordspacing
A.~Fedorov, R.~Beichel, J.~Kalpathy-Cramer, J.~Finet, J.-C. Fillion-Robin, S.~Pujol, C.~Bauer, D.~Jennings, F.~Fennessy, M.~Sonka, J.~Buatti, S.~Aylward, J.~V. Miller, S.~Pieper, and R.~Kikinis, ``3d slicer as an image computing platform for the quantitative imaging network,'' \emph{Magnetic Resonance Imaging}, vol.~30, no.~9, pp. 1323--1341, 2012, quantitative Imaging in Cancer. [Online]. Available: \url{https://www.sciencedirect.com/science/article/pii/S0730725X12001816}
\BIBentrySTDinterwordspacing

\bibitem{Danchilla2012}
\BIBentryALTinterwordspacing
B.~Danchilla, \emph{Three.js Framework}.\hskip 1em plus 0.5em minus 0.4em\relax Berkeley, CA: Apress, 2012, pp. 173--203. [Online]. Available: \url{https://doi.org/10.1007/978-1-4302-3997-0\_7}
\BIBentrySTDinterwordspacing

\bibitem{RII_2023}
\BIBentryALTinterwordspacing
{Research Imaging Institute - UT Health San Antonio}, ``Rii-mango/nifti-reader-js: A javascript nifti file format reader.2,'' Nov 2023. [Online]. Available: \url{https://github.com/rii-mango/NIFTI-Reader-JS}
\BIBentrySTDinterwordspacing

\bibitem{Django_2023}
\BIBentryALTinterwordspacing
Django, ``Django (version 4.2) [computer software],'' Oct 2023. [Online]. Available: \url{https://www.djangoproject.com/}
\BIBentrySTDinterwordspacing

\end{thebibliography}

\vspace{12pt}

\end{document}